\documentclass[twocolumn,prl,10pt,amsmath,amssymb,nofootinbib,showpacs,superscriptaddress,floatfix]{revtex4-1}

\DeclareFontFamily{U}{rcjhbltx}{}
\DeclareFontShape{U}{rcjhbltx}{m}{n}{<->rcjhbltx}{}
\DeclareSymbolFont{hebrewletters}{U}{rcjhbltx}{m}{n}

\newcommand{\rs}{\rm\scriptscriptstyle}

\usepackage{graphicx}
\usepackage{color}
\usepackage[usenames,dvipsnames]{xcolor}
\usepackage[colorlinks=true,linkcolor=Red,citecolor=Green,linktoc=page]{hyperref}
\usepackage{multirow}
\usepackage{float}
\usepackage{flushend}
\usepackage{balance}
\usepackage[varg]{txfonts}
\usepackage{ulem}
\usepackage{fancyhdr}

\DeclareMathSymbol{\lamed}{\mathord}{hebrewletters}{108}

\begin{document}
\title{Quantum magnetic monopole condensate}
\author{	M.\,C.\,Diamantini}
\affiliation{NiPS Laboratory, INFN and Dipartimento di Fisica e Geologia, University of Perugia, via A. Pascoli, I-06100 Perugia, Italy}
\author{C.\,A.\,Trugenberger}
\affiliation{SwissScientific Technologies SA, rue du Rhone 59, CH-1204 Geneva, Switzerland}
\author{V.\,M.\,Vinokur}
\affiliation{Terra Quantum AG, St. Gallerstrasse 16A, CH-9400 Rorschach, Switzerland}

\begin{abstract}
Despite decades-long efforts, magnetic monopoles were never found as elementary particles. Monopoles and associated currents were directly measured in experiments and identified as topological quasiparticle excitations in emergent condensed matter systems. These monopoles and the related electric-magnetic symmetry were restricted to classical electrodynamics, with monopoles behaving as classical particles. Here we show that the electric-magnetic symmetry is most fundamental and extends to full quantum behavior. We demonstrate that at low temperatures magnetic monopoles can form a quantum Bose condensate dual to the charge Cooper pair condensate in superconductors. The monopole Bose condensate manifests as a superinsulating state with infinite resistance, dual to superconductivity. Monopole supercurrents result in the electric analog of the Meissner effect and lead to linear confinement of Cooper pairs by Polyakov electric strings  in analogy to quarks in hadrons.
\end{abstract}

\maketitle


\section*{Introduction}~~\\
Maxwell's equations in vacuum are symmetric under the duality transformation ${\bf E} \to {\bf B}$ and ${\bf B} \to -{\bf E}$ (we use natural units $c=1$, $\hbar = 1$, $\varepsilon_0=0$). Duality is preserved, provided that both electric and magnetic sources (magnetic monopoles and magnetic currents) are included\,\cite{olive}.
Magnetic monopoles, 
while elusive as elementary particles\,\cite{milton}, exist in many materials in the form of emergent quasiparticle excitations\,\cite{qi}. Magnetic monopoles and associated currents were directly measured in experiments\,\cite{zeldov}, confirming the predicted symmetry between electricity and magnetism.
The existence of monopoles requires that gauge fields are compact, implying, in turn, the quantization of charge and Dirac strings\,\cite{dirac1931} or a core with additional degrees of freedom to regularize the singularities of the vector potential\,\cite{thooft1974, polyakov1974}. 

The importance of electric-magnetic duality was first realized by Nambu\,\cite{nambu}, Mandelstam\,\cite{mandelstam} and 't Hooft\,\cite{thooft}, who proposed that colour confinement in quantum chromodynamics (QCD) can be understood as a dual Meissner effect. In the present context, electric-magnetic duality manifestly realizes the symmetry between Cooper pairs, which are Noether charges, and magnetic monopoles, which are topological solitons, and forms the foundation for the superconductor-insulator transition\,\cite{Blanter1997} and the appearance of the superinsulating state\,\cite{dst,vinokur,dtv1}. Magnetic monopoles arise also as instantons in Josephson junction arrays (JJA)\,\cite{dtnpv}, which are easily accessible experimental systems themselves and provide an exemplary model for superconducting films\,\cite{Tinkham}.

So far, monopoles in emergent condensed matter systems were treated as classical excitations.
Here we show that the electric-magnetic symmetry is most fundamental and extends to the full quantum realm. 
We demonstrate here that at low temperatures, magnetic monopoles form a quantum Bose condensate dual to the charge condensate in superconductors and generate a superinsulating state with infinite resistance, dual to superconductivity\,\cite{dst, vinokur}. We show that magnetic monopole supercurrents result in the direct electric analog of the Meissner effect and lead to linear confinement of the Cooper pairs by Polyakov's electric strings\,\cite{mironov, dtv1, dtv2} (dual to superconducting vortices) in analogy to quarks within hadrons\,\cite {greensite}.
The monopole condensate realizes a 3D version of superinsulators, that have been previously observed in 2D superconducting films, and result from quantum tunneling events, or instantons\,\cite{dst, vinokur, mironov, dtv1, dtv2}. 

\begin{figure}[t!]
	\includegraphics[width=9cm]{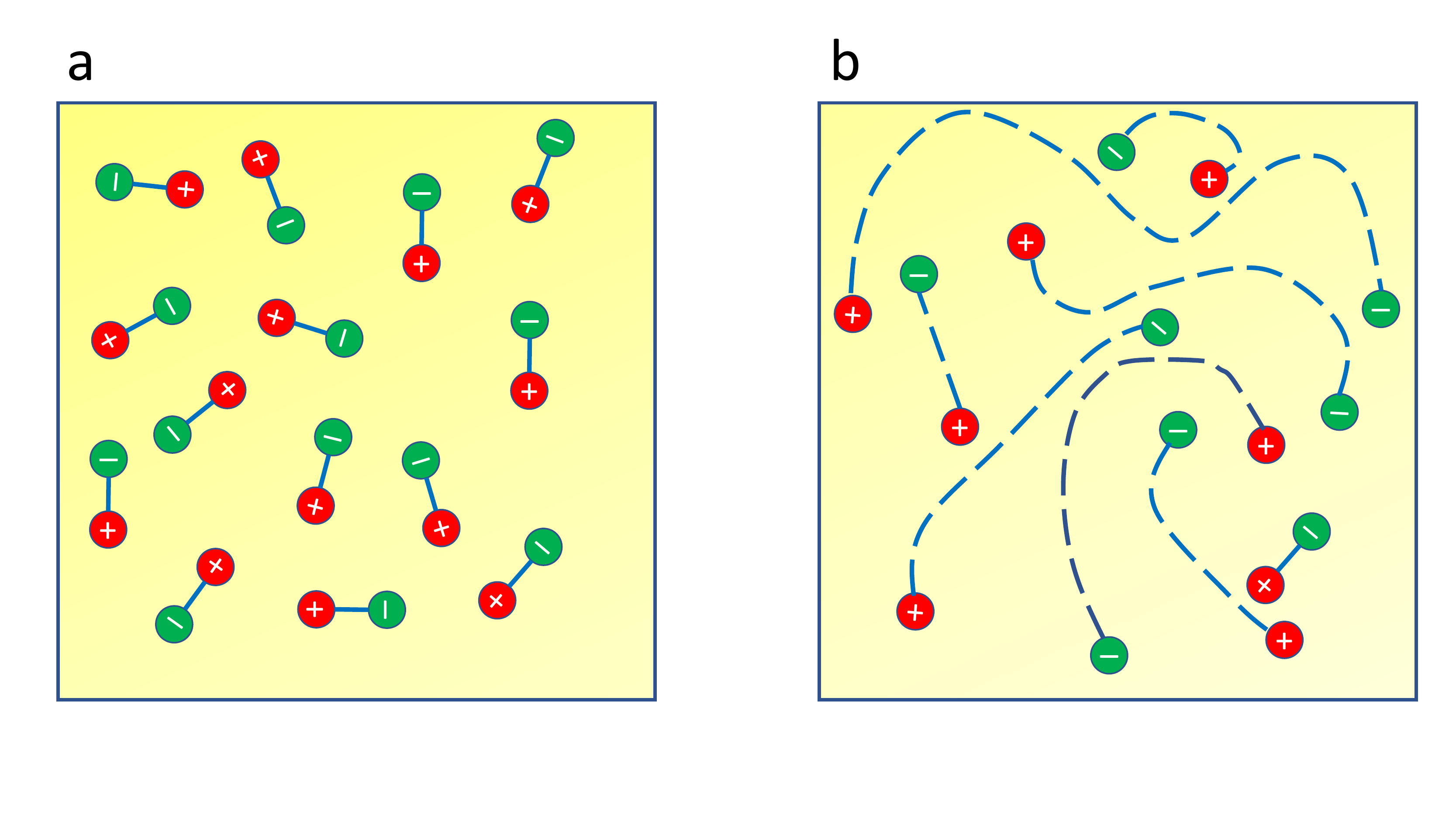}
	\vspace{-0.5cm}
	\caption{\textbf{Magnetic monopole states at low temperatures.} \textbf{a:} Monopoles are confined into dipoles state by tension of vortices connecting them. \textbf{b:} As the vortex tension vanishes, the monopoles at their endpoints become loose free particles that can condense.} 
	\label{Fig1}
\end{figure}

\section*{Results}
\subsection{Magnetic monopoles in granular superconductors}~~\\
To gain insight into the nature of a system that can harbor Cooper pairs and magnetic monopoles simultaneously, we first reiterate that monopoles naturally emerge in 2D JJA\,\cite{dtnpv} as instantons and provide the underlying mechanism of 2D superinsulation as quantum tunneling events. To establish that JJA indeed offers a universal model describing superinsulation in films, one recalls the early hypothesis that, in the vicinity of the superconductor-insulator transition (SIT), the films acquire self-induced electronic granular texture and can be viewed as a set of superconducting granules immersed into an insulating matrix\,\cite{Kowal1994}. Employing the model of the 2D JJA\,\cite{Fistul2008} to treat the experimental data of\,\cite{Vinokur2013,mironov}, perfectly confirmed this picture, and settled that granules have the  typical size of order of the superconducting coherence length, $\xi$, and are coupled by Josephson links. On the theory side, this granular structure was derived in the framework of the gauge theory of the SIT in\,\cite{dtv2019}, conclusively establishing the JJA-like texture in 2D systems experiencing the SIT. 

As a next step, we generalize the reasoning of\,\cite{dtv2019} onto 3D systems. This enables us to adopt the original gauge theory of JJA\,\cite{dst} for 3D superconductors and consider a general inhomogeneous system of superconducting granules, i.e. bubbles of Cooper pair condensate, coupled by tunneling links. Then, depending on the ratio of the strength of Josephson coupling and the Coulomb energy of the elemental excessive charge on a single granule, the system can be either an insulator, superconductor, or topological insulator\,\cite{dst,dtv2019}.
If the system is at the insulating side of the SIT, the global phase coherence is absent and each condensate bubble is characterized by an independent phase. The relevant degrees of freedom in such systems are single Cooper pairs that can tunnel from one island to the next one, leaving behind a Cooper hole, and vortices, resulting from non-trivial phase circulations over adjacent granules. In a 2D system such a vortex would be a usual Josephson vortex, and in 3D such an elemental 'minimal' vortex is similar to a pancake vortex in a layered cuprate\,\cite{pancake1}. In the usual configuration, these pancake vortices aggregate on top of each other to make stacks, or chains. When the pancake stack forms, the magnetic monopoles and antimonopoles at the ``bottom" and ``top" of each of such a pancake ``annihilate" and one long vortex forms. Only the monopoles and anti-monopoles at the very end of this configuration survive and, correspondingly, one has such monopoles and anti-monopoles only at the surfaces of the sample\,\cite{pancake1}. When pancake vortices are ballistic and the layers are only weakly coupled, however, the vertical stacks can break in the middle, resulting in monopole and anti-monopole pairs joined by a shorter vortex in the interior of the sample. As always, when there are both dynamic charges and vortices in the spectrum, the latter acquires a topological gap\,\cite{wilczek}. The inverse of the gap sets the spacial scale for the width of the vortices and radius of monopoles. 

On distances larger than the vortex width, long vortices appear as quantized flux tube singularities that become unobservable because of the compactness of the U(1) gauge fields\,\cite{olive}, and a neutral plasma of point magnetic monopoles satisfying the Dirac quantization\,\cite{olive} condition forms.  Remarkably, as we show below, the monopole plasma strongly suppresses Cooper pair tunneling providing thus a mechanism that stabilizes the granular structure. 
\\

\subsection{Long-distance effective field theory}~~\\
Having established that in the vicinity of the SIT an ensemble of Cooper pairs acquires a granular structure, we construct a Ginzburg-Landau-type effective field theory of such granular Cooper pair condensates. We focus on the London limit, i.e. on long distances, much larger than the topological width scale so that vortices and monopoles appear as point-like singularities. As a first step, following general principles by Wilczek\,\cite{wilczek}, we identify that the dominant interactions are the topological mutual statistics interaction setting the Aharonov-Bohm phases between charges and vortices. A standard description of the interaction part of the system's Lagrangian is achieved by introducing  two fictitious gauge fields, a vector field  $a_{\mu}$ and an antisymmetric pseudotensor gauge field $b_{\mu \nu}$\,\cite{blau}:
\begin{equation}
	{\cal L}_{\rm BI} = {1\over 4\pi} b_{\mu \nu} \epsilon^{\mu \nu \alpha \beta} \partial_{\alpha} a_{\beta}  + \ a_{\mu} q^{\mu} +{1\over 2} b_{\mu \nu} m^{\mu \nu} \ .
	\label{bosonictopins}
\end{equation}
where $q^{\mu}$ and $m^{\mu \nu}$ are the charge and vortex currents, respectively. 
This so-called BF model\,\cite{blau} is topological, since it is metric-independent. It is invariant under the usual gauge transformations  $a_{\mu} \to a_{\mu} + \partial_{\mu} \xi$ and under the gauge transformation of the second kind\cite{blau}, under which the antisymmetric tensor transforms as $b_{\mu \nu} \to b_{\mu \nu} +\partial_{\mu} \lambda_{\nu} -\partial_{\nu} \lambda_{\mu}$, a vector field $\lambda_{\mu}$ becoming itself the gauge function. The BF action for a model defined on a compact space endowed with the non-trivial topology, yields a ground state with the degeneracy reflecting, one-to-one, this topology and is referred to as topological order\,\cite{wen2013}. The coefficient $1/4\pi$ of the first term in Eq.\,(\ref{bosonictopins}) ensures that the system does not have such a topological order\,\cite{semenoff}. In turn, the topological coupling between a vector and a pseudotensor in 3D, ensures the parity (${\mathbb Z}^{\rs P}$) and time-reversal (${\mathbb Z}^{\rs T}$) symmetries of the model. The field strength associated with $a_{\mu}$ is  $f_{\mu \nu}$$=$$\partial_{\mu} a_{\nu} -\partial_{\nu} a_{\mu}$, whereas $b_{\mu \nu}$ has the associated 3-tensor field strength $h_{\mu \nu \alpha}$$=$$\partial_{\mu} b_{\nu \alpha}  + \partial_{\nu} b_{\alpha \mu} + \partial_{\alpha} b_{\mu \nu}$. It can be easily checked that this expression is invariant under the gauge transformations of the second kind; it plays for $b_{\mu \nu}$ the same role that the field strength $f_{\mu \nu}$ plays for the usual gauge field $a_{\mu}$. 
The dual field strengths $j^{\mu}$$=$$(1/ 2\pi) h^{\mu}$$=$$(1/4\pi) \epsilon^{\mu \nu \alpha \beta} \partial_{\nu} b_{\alpha \beta} $ and $m^{\mu \nu} = (1/2\pi) \tilde f^{\mu \nu} =(1/4\pi) \epsilon^{\mu \nu \alpha \beta} f_{\mu \nu} $ represent the conserved charge and vortex currents, respectively. For Cooper pairs, the charges are measured in integer units of the elemental charge of a Cooper pair, $2e$. Accordingly, the magnetic charge of  monopoles and vortices is measured in integer units of $2\pi / 2e = \pi / e$. If one defines a fundamental charge as a charge of a single electron, $e$, then our magnetic monopoles of strength $\pi/e$ should be viewed, strictly speaking, as half-monopoles. However, we always deal with the phases of matter where charge unit is $2e$ rather than $e$. Thus  we shall call our objects the unit monopoles. This complies with condensed matter notations, where a unit vortex carries magnetic flux quantum $\pi/e$, since it always appears in a Cooper pair condensate, but differs from the field theory notation where $\pi/e$ is half a vortex. 

The developed model describes what is known today as a (simple, as opposed to strong) bosonic topological insulator\,\cite{senthil} and is an intensely investigated state of matter. Open vortices emerging in this model carry magnetic monopole-antimonopole pairs with current $m^{\nu } = \partial_{\mu} m^{\mu \nu}$ at their ends, and the gauge symmetry of the second kind is broken. 
The state of the monopole ensemble is self-consistently harnessed with vortex properties. Existence of the appreciable vortex tension implies that vortices are short and taut and linearly confine the monopoles into `small' dipoles, as illustrated in Fig.\,1a. The vanishing vortex tension allows monopoles to break loose, while vortices themselves grow infinitely long and loose and turn into unobservable Dirac strings, as shown in Fig.\,1b. 
Accordingly, the world-lines of monopoles become infinitely long as well, which means that they Bose condense. Determining the condensation point is a dynamical problem. One can say that the condensation point is set by the moment when quantum corrections to the vortex tension are large enough to turn it negative so that vortices become Dirac strings. The rest of this paper is devoted to answering this question and to deriving the nature of the ensuing new state of matter. 

To conclude here, we generalize the 2D consideration of\,\cite{dst, dtv2019} onto 3D systems and predict that in three dimensions superconductors also acquire self-induced emergent granularity in the vicinity of the SIT. Magnetic monopoles play a crucial role in the formation and properties of this new superconducting state.

\section*{Phase transitions and phase diagram}~~\\
Let us consider a granular system (irrespective to whether the granularity is the self-induced electronic granularity\,\cite{vinokur} or is of the structural origin, such as, e.g. granular diamond\,\cite{moschalkov}), characterized by the length scale $\ell$ playing the role of the granule size, and examine the various phases that can emerge. We focus on cubic lattices since, as in 2D, the paradigmatic system for monopoles and the transitions they induce is a Josephson junction array\,\cite{jja}. Of course, the universality class of the transition can depend on the lattice details but the discussion of such effects is beyond the scope of the present work. We thus formulate the action (\ref{bosonictopins}) on a cubic lattice of spacing $\ell$ and we add all possible local gauge invariant terms. In the presence of magnetic monopoles the tensor current $m^{\mu \nu}$ does not conserve any more, and the gauge invariance of the second kind of the tensor field $b_{\mu \nu}$ is effectively broken at the vortex endpoints. Longitudinal components of the tensor gauge field $b_{\mu \nu}$ become the usual vector gauge fields for the magnetic monopoles and induce for the monopoles the same type of Coulomb interaction which experience electric charges. However, this Coulomb interaction is subdominant to the linear tension created by vortices connecting monopole and anti-monopole pair, and one can neglect it when determining the phase structure. More specifically, when inverting the Kalb-Ramond kernel, we will consider only the transverse components of the vortices and neglect their endpoints.

Rotating to Euclidean space-time we arrive at the action
\begin{eqnarray}
	S = \sum_x {\ell ^4\over 4f^2}
	f_{\mu \nu} f_{\mu \nu} -i{\ell ^4\over 4\pi }
	a_{\mu }k_{\mu \alpha \beta }b_{\alpha \beta } + {\ell ^4\over 12 g^2}
	h_{\mu \nu \alpha }h_{\mu \nu \alpha} 
	\nonumber \\
	+i\ell a_{\mu }q_{\mu }
	+i\ell ^2{1  \over 2} b_{\mu \nu}m_{\mu \nu} \ ,
	\label{fullaction}
\end{eqnarray}
where $k_{\mu \nu \alpha}$ is the lattice BF term\,\cite{dst}, see Methods, $f$ is a dimensionless coupling, and $g$ has the canonical dimension of mass ([mass]). To describe materials with the relative electric permittivity $\varepsilon$ and relative magnetic permeability $\mu$, we incorporate the velocity of light $v=1/\sqrt{\varepsilon \mu} <1 $ by defining the Euclidean time lattice spacing as $\ell_0 = \ell/v$ and by rescaling all time derivatives, currents, and zero-components of gauge fields by the factor $1/v$.  As a consequence, both gauge fields acquire a dispersion relation $E = \sqrt{m^2 v^4 + v^2 {\bf p}^2}$ with the topological mass\,\cite{bowick} $m= fg/2\pi v$. The dimensionless parameter $f={\cal O}(e)$ encodes the effective Coulomb interaction strength in the material, $g$ is the magnetic scale $g={\cal O}(1/\lambda_{\rs L})$, where $\lambda_{\rs L}$ is the London penetration depth of the superconducting phase. 

To analyze how the additional interactions can drive quantum phase transitions taking the system out of the bosonic insulator, we integrate over the fictitious gauge fields to obtain a Euclidean action $S_{\rm cv}$ for point charges and line vortices alone. As we show in Methods, this is proportional to the length of the charge world-lines and to the area of vortex world-surfaces, exactly as their configurational entropy. Charges and vortices can thus be assigned an effective action (equivalent to a quantum ``free energy" in this Euclidean field theory context)
\begin{equation}
	F_{\rm cv} = \left( s_\mathrm{c} - h_\mathrm{c} \right) N + \left( s_\mathrm{v} - h_\mathrm{v} \right) A \ ,
	\label{free}
\end{equation}
where $N$ and $A$ are the length of word-lines in number of lattice links and the area of world-surfaces in number of lattice plaquettes, respectively and $s_\mathrm{c,v}$ and $h_\mathrm{c,v}$ denote the action and entropy contributions (per length and area) of charges and vortices, respectively. The possible phases that are realized are determined by the relation between the values of the Coulomb and magnetic scales and by materials parameters determining whether the coefficients of $N$ and $A$ are positive or negative. For positive coefficients, long world-lines and large wold-surfaces are suppressed. If either of the parenthesis becomes negative, then either a charge or a monopole condensate forms. In the case of charges, the proliferation of long-world lines is the geometric picture of Bose condensation first put forth by Onsager\,\cite{onsager1} and elaborated by Feynman\,\cite{feynman}, see\,\cite{schakel} for a recent discussion. In the case of vortices, the string between magnetic monopole endpoints becomes loose and assumes the role of an unobservable Dirac string. In this case, the monopoles are characterized by long world-lines describing their Bose condensate phase, see Fig.1. The details of this vortex transition have been discussed in\,\cite{kleinert,geostring}. The resulting phase diagram is determined by the value of the parameter $\eta=2\pi (mv\ell) G/\sqrt{\mu_{\rs N} \mu_{\rs A}}$ encoding the strength of quantum fluctuations and the material properties of the system\,\cite{dtv2} and tuning parameter, $\gamma= (f/\ell g) \sqrt{\mu_{\rs N}/\mu_{\rs A}}$, taking the system across superconductor-insulator transition (SIT). Here $G={\cal O}(G(mv\ell))$, where $G(mv\ell)$ is the diagonal element of the lattice kernel $G(x-y)$ representing the inverse of the operator $\ell^2 \left( (mv)^2 -\nabla^2 \right) $, and $\mu_{\rs N}$ and $\mu_{\rs A}$ are the entropy per unit length of the world line and per unit area of the world surface, respectively. The phase structure at $T=0$ and the domains of different phases in the critical vicinity of the SIT are defined by the relations, see Methods for details:
\begin{eqnarray}
	\eta < 1 \to 
	\begin{cases}
		\gamma< 1 \ , {\rm charge \ Bose \ condensate} \ ,\\
		\gamma> 1\ , {\rm monopole\ Bose\ condensate} \ ,\\
	\end{cases}
	\nonumber \\
	\eta> 1 \to
	\begin{cases}
		\gamma < {1\over \eta}
		\ , {\rm charge\ Bose\ condensate} \ ,\\
		{1\over \eta}
		<\gamma<\eta \ , {\rm bosonic\ insulator} \ ,\\
		\gamma > \eta \ , {\rm monopole \ Bose \ condensate }\ .\\
	\end{cases}
	\label{phasestructure}
\end{eqnarray}
%
and are shown in Fig.\,2. The finite-temperature decay of the condensates, corresponding to the deconfinement transitions into the bosonic insulator, is described by the same approach\,\cite{diamantini2020}. 

The effect of charge condensation is immediately revealed by minimally coupling the current $j^{\mu} = (1/2\pi) h^{\mu} $ to the electromagnetic gauge potential $A_{\mu}$ and integrating out all matter fields to obtain the electromagnetic response. The latter acquires a photon mass term $\propto A_{\mu} A^{\mu}$ so that the induced charge current $j_{\mu} \propto A_{\mu}$. This is nothing but London equation i.e. the major manifestation of superconductivity\,\cite{Tinkham}. Not dwelling on this standard derivation, but just stressing that our approach reproduces that charge condensate phase is a superconductor, we now focus on the new phase induced by the condensation of magnetic monopoles. 
At $f/\ell < {\cal O}(1) $, a superconducting phase is thus realized, as observed in granular diamond\,\cite{moschalkov}. 
For $\ell g/f< {\cal O}(1)$, however, there is a dual superinsulating phase governed by the magnetic monopole condensate, which is the main subject of this paper. 

To conclude this section, we would like to point out that the direct transition between superinsulator and superconductor for $\eta <1$ shown in Fig. 2 matches perfectly the low-temperature, quantum phase structure of the QCD as a function of the density, with a transition between confined hadronic matter and colour superconductivity\,\cite{qcd}. 

\section*{Electromagnetic response and the electric string tension}~~\\
To reveal the nature of the superinsulating phase, we examine its electromagnetic response. To that end we again minimally couple the electric current $j_{\mu}$ to the electromagnetic gauge field $A_{\mu}$ and compute its effective action, see Methods. Taking the limit $mv\ell \gg1$, we find 
\begin{eqnarray}
	{\rm exp} \left( -S_{\rm e.m.} \right) = \sum_{t_{\mu \nu}} {\rm e}^{ - {1\over 8 f^2} \sum_{x, \mu, \nu} \left( F_{\mu \nu} -2\pi t_{\mu \nu} \right)^2}  \, .
	\label{villain}
\end{eqnarray}
The monopole condensation for strong $f$ renders the real electromagnetic field a compact variable, defined on the interval $[-\pi, +\pi]$ and the electromagnetic response is given by Polyakov's compact QED action\,\cite{polyakovoriginal, polyakov}. 
This changes drastically the Coulomb interaction. To see that, let us take two external probe charges $\pm q_{\rm ext}$ and find the expectation value for the corresponding Wilson loop operator $W(C)$, where $C$ is the closed loop in 4D Euclidean space-time (the factor $\ell$ is absorbed into the gauge field $A_{\mu}$ to make it dimensionless)
\begin{equation}
	\langle W(C) \rangle = {1\over Z_{A_{\mu}, t_{\mu \nu}}} \sum_{\{ t_{\mu \nu} \} }  \int_{-\pi}^{+\pi}  {\cal D} A_{\mu} 
	\ {\rm e}^{-{1\over 8 f^2}  \ \sum_{x,\mu \nu} \left( F_{\mu \nu} -2\pi t_{\mu \nu} \right)^2 } {\rm e}^{iq_{\rm ext} \sum_C l_{\mu} A_{\mu} } \ ,
	\label{wilson1}
\end{equation}
where $l_{\mu}$ takes the value 1 on the links forming the Wilson loop $C$ and 0 otherwise. 
When the loop $C$ is restricted to the plane formed by the Euclidean time and one of the space coordinates, $\langle W(C)\rangle$ measures the interaction energy between charges $\pm q_{\rm ext}$.  A perimeter law indicates a short-range potential, while an area-law is tantamount to a linear interaction between them\,\cite{polyakov}. 
For Cooper pairs, $q_{\rm ext}= 1$, see Methods, $\langle W(C) \rangle = {\rm exp}({- \sigma A})$ 
where $A$ is the area of the surface $S$ enclosed by the loop $C$. This yields a linear interaction between probe Cooper pairs, which therefore can be viewed as confined by the elastic string with the string tension
\begin{equation}
	\sigma ={32 f\over \pi \sqrt{\varepsilon \mu}} {1\over \ell^2} {\rm exp}\left({-{\pi G(0) \over 8 f^2}}\right) \ \ ,
	\label{arealaw}
\end{equation}
where $G(0)$$=$$0.155$ is the value of the 4D lattice Coulomb potential at coinciding points. The monopole condensate, thus, generates a string binding together charges and preventing charge transport in systems of a sufficient spatial size. A magnetic monopole condensate is a 3D superinsulator, characterized by an infinite resistance at finite temperatures\,\cite{dst, vinokur, mironov, dtv1, dtv2}. The critical value of the effective Coulomb interaction strength for the transition to the superinsulating phase is $f_{\rm crit} = {\cal O}( \ell/ \lambda)$. 
\begin{figure}[t!]
	\includegraphics[width=7.7cm]{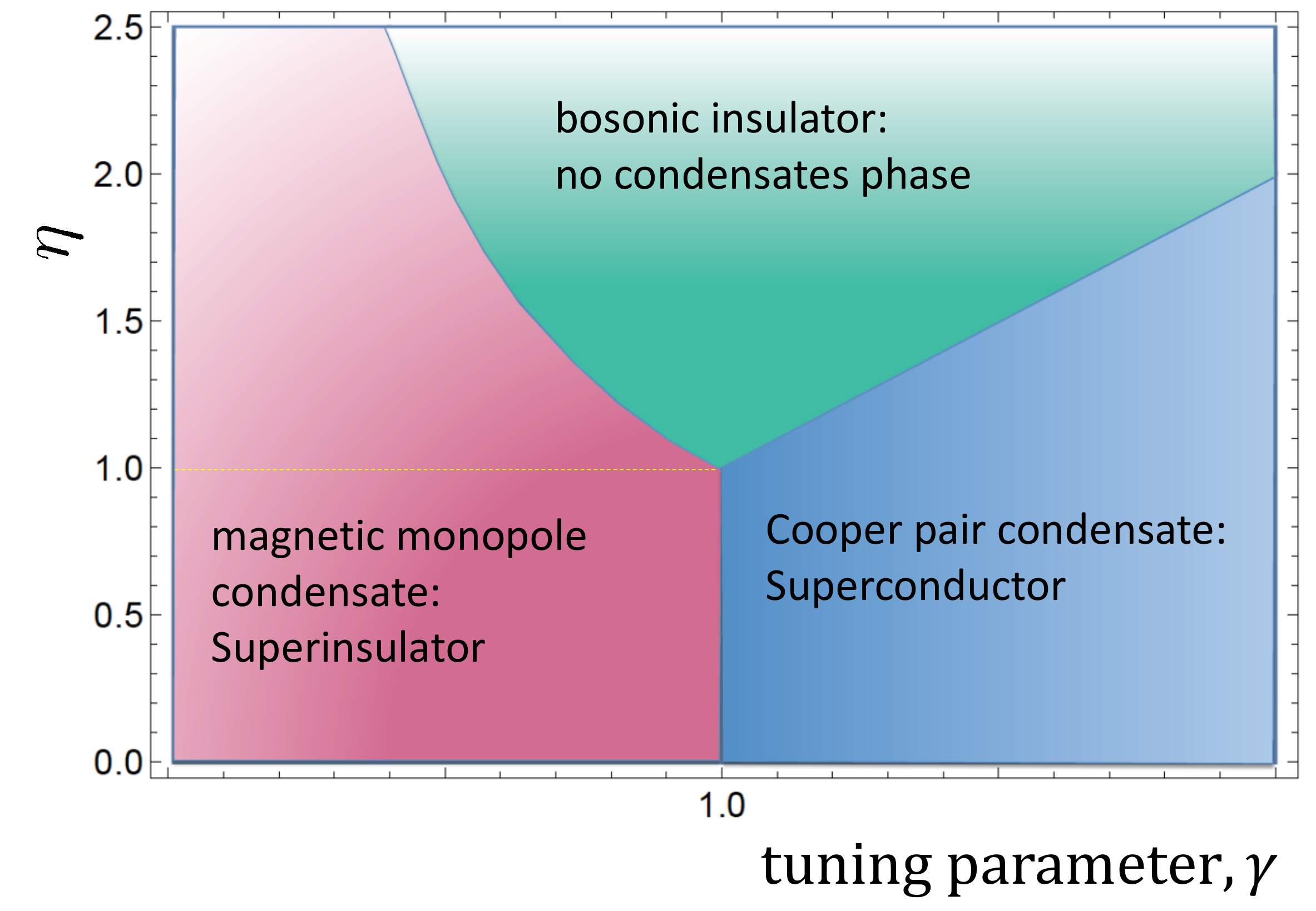}
	\vspace{-0.2cm}
	\caption{\textbf{Cooper pairs--quantum magnetic monopoles phase diagram at $T$$=$$0$.} The parameter $\eta=2\pi (mv\ell) G/\sqrt{\mu_{\rs N} \mu_{\rs A}}$ encodes the strength of quantum fluctuations and the material properties of the system. The inter-phases lines are $\eta$$=$$1/\gamma~(\gamma$$<$$1,\eta$$>1)$, $\gamma$$=$$1$~ ($\eta$$<$$1$), and $\eta$=$\gamma$~($\eta$$>$$1$, $\gamma$$>$$1$) .   }
	\label{Fig1}
\end{figure}

\section*{Discussion}~~\\
Superinsulation has been observed in 2D, where magnetic monopoles are instantons rather than particles\,\cite{vinokur, mironov, dtv2} and the structure of the phase diagram is conclusively established. The charge Berezinskii-Kosterlitz-Thouless (BKT) transition into the low-temperature confined superinsulating phase was measured.
Reference\,\cite{dtv2} presented measurements of the electromagnetic response of the superinsulating phase and of properties of electric strings. In particular, the observed double kinks in the $I$-$V$ characteristics indicate the predicted electric Meissner effect characteristic to superinsulators. The experimental results exhibit a fair agreement with the theoretical predictions. The existence of a bosonic insulator in various materials and, in particular, in the same NbTiN films where the kinked $I$-$V$ curves were measured, was unambiguously established in\,\cite{dtv4}.  That TiN films with smaller $\eta<1$ exhibit the direct SIT, while NbTiN films endowed with $\eta>1$ show the SIT across the intermediate topological insulator phase, complies with the theoretical expectations. However, more research on various systems is required to spot the exact location of the tricritical point.  To establish the Cooper pair-based nature of the bosonic topological insulator, shot noise measurements similar to those carried out in\,\cite{Zhou2019} are desirable. 
The signature of 3D superinsulation\,\cite{dtv3} has been detected in InO films\,\cite{shahar2005,ovadia2015}. However, more studies are necessary for drawing reliable conclusions that InO can be considered as a material hosting a magnetic monopole condensate. Strongly type II superconductors with a fine, inherent or self-induced granular electronic texture, are other most plausible candidates to house 3D superinsulators. Finally, another class of candidates are layered materials. Vortex lines in such materials can be regarded as stacks of pancake vortices\,\cite{pancake1}. If the layers are weakly coupled and vortices are ballistic, the pancakes can split and form magnetic monopoles at the inner layers' intersections, albeit very anisotropic ones. These can then condense into a superinsulating phase.

Finally,  while the monopole condensate existence is strongly supported by the observation of the superinsulating state and the corresponding experimental implications are reliably established by recent transport measurements\,\cite{dtv2}, the conclusive evidence for monopoles should come from their direct observations. One of the ways to implement such an observation may be extending the SQUID-on-tip device method of\,\cite{zeldov} to lower temperatures.

\bigskip

\section*{Data availability}~~\\
Data sharing not applicable to this article as no datasets were generated or analyzed during the current study.

\medskip
\textbf{Acknowledgements} \par 
V.M.V. thanks Terra Quantum for support at the final stage of the work.
M.C.D. thanks CERN, where she completed this work, for kind hospitality.

\section*{Appendix}
\subsection{Lattice BF term}
To formulate the gauge-invariant lattice $BF$-term, we follow\,\cite{dst} and introduce the lattice $BF$ operators
\begin{eqnarray}
	k_{\mu \nu \rho} \equiv S_{\mu }\epsilon _{\mu \alpha
		\nu \rho } d _{\alpha } \ ,
	\nonumber \\
	\hat k_{\mu \nu \rho} \equiv \epsilon _{\mu \nu \alpha \rho}
	\hat d_{\alpha }\hat S_{\rho } \ ,
	\label{kop}
\end{eqnarray}
where 
\begin{eqnarray}
	d_{\mu } f(x) \equiv {{f(x+\ell \hat \mu )-f(x)}\over
		\ell}\ ,\ \  S_{\mu }f(x) \equiv f(x+\ell \hat \mu )\ ,
	\nonumber \\
	\hat d_{\mu } f(x) \equiv {{f(x)-f(x-\ell \hat \mu )}\over \ell } \ ,
	\ \ \hat S_{\mu }f(x) \equiv f(x-\ell \hat \mu ) \ ,
	\label{shift}
\end{eqnarray}
are the forward and backward lattice derivative and shift operators, respectively. Summation by parts on the lattice interchanges both the two derivatives (with a minus sign) and the two shift operators; gauge transformations are defined using the forward lattice derivative. The 
two lattice $BF$ operators are interchanged (no minus sign) upon summation by parts on the lattice and are gauge invariant so that:
\begin{eqnarray}
	k_{\mu \nu \rho} d _{\nu } = k_{\mu \nu \rho}
	d _{\rho } = \hat d_{\mu } k_{\mu \nu \rho} = 0 \ ,
	\nonumber \\
	\hat k_{\mu \nu \rho }d _{\rho } = \hat d _{\mu }
	\hat k_{\mu \nu \rho} = \hat d_{\nu }
	\hat k_{\mu \nu \rho } = 0 \ .
	\label{gaugeinv}
\end{eqnarray}
And satisfy the equations
\begin{eqnarray}
	\hat k_{\mu \nu \rho} k_{\rho \lambda \omega} =
	-\left( \delta _{\mu \lambda} \delta_{\nu \omega} - \delta _{\mu \omega}
	\delta_{\nu \lambda } \right) \nabla^2  
	\nonumber \\
	+ \left( \delta _{\mu \lambda }
	d_{\nu } \hat d_{\omega} - \delta _{\nu \lambda } d_{\mu }
	\hat d_{\omega } \right) 
	+ \left( \delta _{\nu \omega} d_{\mu }
	\hat d_{\lambda } - \delta _{\mu \omega} d_{\nu } \hat
	d_{\lambda } \right) \ ,
	\nonumber \\
	\hat k_{\mu \nu \rho} k_{\rho \nu \omega } = k_{\mu \nu \rho } \hat
	k_{\rho \nu \omega} = 2 \left( \delta _{\mu \omega } \nabla^2 - d_{\mu }
	\hat d_{\omega } \right) \ ,
	\label{maxwell}
\end{eqnarray}
where $\nabla^2 = \hat d_{\mu} d_{\mu}$ is the lattice Laplacian. 
We use the notation $\Delta_{\mu}$ and $\hat \Delta_{\mu}$ for the forward and backwards finite difference operators. 
\\

\subsection{Phases of monopoles} 
To find the topological action for monopoles, we start from eq. (\ref{fullaction}) and integrate out fictitious gauge fields $a_{\mu}$ and $b_{\mu \nu}$
\begin{eqnarray}
	S_{\rm top} = \sum_x {f^2 \over 2\ell^2} \ q_{\mu }{\delta _{\mu \nu }
		\over {(mv)^2- \nabla ^2 }} q_{\nu } 
	+ {g^2 \over 8} \ m_{\mu \nu}
	{{\delta _{\mu \alpha } \delta _{\nu \beta }-\delta _{\mu \beta}
			\delta _{\nu \alpha }} \over {(mv)^2-\nabla ^2}} m_{\alpha \beta } 
	\nonumber \\
	+i {\pi (mv)^2 \over \ell}
	\ q_{\mu } {k_{\mu \alpha \beta}
		\over {\nabla ^2 \left( (mv)^2- \nabla ^2 \right) }} m_{\alpha \beta } \ ,
	\label{topeffac}
\end{eqnarray}
The last term represents the Aharonov-Bohm phases of charged particles around vortices of width $\lambda_{\rs L}$. On scales much larger than $\lambda_{\rs L}$, where the denominator reduces to $(mv)^2 \nabla^2$, this term becomes $i2\pi -{\rm integer}$, as can be easily recognized by expressing $q_{\mu} = (1/2) \ell k_{\mu \alpha \beta} y_{\alpha \beta}$. This reflects the absence of Aharonov-Bohm phases between charges $ne$ and magnetic fluxes $2\pi/ne$. Accordingly, we shall henceforth neglect this term. 

The important consequence of the topological interactions is that they induce self-energies in form of the mass of Cooper pairs and tension for vortices between magnetic monopoles. These self-energies are encoded in the short-range kernels in the action (\ref{topeffac}), which we approximate by a constant. World-lines and world-surfaces 
are thus assigned ``energies" (formally Euclidean actions in the present statistical field theory setting and thus dimensionless in our units) proportional to their length $N$ and area $A$ (measured in numbers of links and plaquettes),
\begin{eqnarray}
	S_{\rs N} =2 \pi (mv\ell) G \ {f\over g\ell} \ Q^2 N \ ,
	\nonumber \\
	S_{\rs A} = 2\pi (mv\ell) G \ {g\ell \over f} \  M^2 A \ ,
	\label{energies}
\end{eqnarray}
where $G=O(G(mv\ell))$, with $G(mv\ell)$ the diagonal element of the lattice kernel $G(x-y)$ representing the inverse of the operator $\ell^2 \left( (mv)^2 -\nabla^2 \right) $, and $Q$ and $M$ are the integer quantum numbers carried by the two kinds of topological defects. However, also the entropy of link strings and plaquette surfaces is proportional to their length and area \cite{nelson}, $ \mu_{\rs N} N$ and $\mu_{\rs A} A$. Both coefficients $\mu$ are non-universal: $\mu_{\rs N} \simeq {\rm ln}(7)$ since at each step the non-backtracking string can choose among 7 possible directions on how to continue, while $\mu_{\rs A}$ does not have such a simple interpretation but can be estimated numerically. This gives for both types of topological defects a ``free energy" proportional to their dimension and with coefficients that can be positive or negative depending on the parameters of the theory. The total free energy is 
\begin{equation}
	{F \over 2\pi (mv\ell) G} = \left[ \left( {f\over g\ell} \ Q^2 -{1\over \eta_{\rs Q}} \right) N 
	\nonumber \\
	+ \left( {g\ell \over f} \  M^2 -{1\over \eta_{\rs M}}\right) A \right] \ ,
	\label{freeenergy}
\end{equation} 
where we have defined 
\begin{equation}
	\eta_{\rs Q}=  {2\pi (mv\ell) G\over \mu_{\rs N}} \ , \qquad \eta_{\rs M}=  {2\pi (mv\ell) G \over \mu_{\rs A}} \ .
	\label{eta}
\end{equation}
If the coefficients are positive, the self-energy dominates and large string/surface configurations are suppressed in the partition function. In this regime Cooper pairs and/or vortices are gapped excitations, suppressed by their large action. 
If the coefficients, instead are negative, the entropy dominates and large configurations are favoured in the ``free energy" (effective action). The phase in which long world-lines of Cooper pairs dominate the Euclidean partition function is a charge Bose condensate, as discussed originally by Onsager\,\cite{onsager1} and Feynmann\,\cite{feynman} (for a recent discussion see\,\cite{schakel}). This phase is the Bose condensate of magnetic monopoles.  For vortices, proliferation of large world-surfaces means that the strings binding monopoles and antimonopoles into neutral pairs become loose. We will show below that in this case the long real monopole world-lines dominate the electromagnetic response. 

The combined energy-entropy balance equations are best viewed as defining the interior of an ellipse on a 2D integer lattice of electric and magnetic quantum numbers, 
\begin{equation}
	{Q^2 \over r_{\rs Q}^2} + {M^2 \over r_{\rs M}^2} < 1 \ ,
	\label{ellipse}
\end{equation}
where the semiaxes are given by
\begin{eqnarray}
	r_{\rs Q}^2  = {\ell g \over f} \ {1\over \eta_{\rs Q}} = {\ell g \over f} \sqrt{\mu_{\rs N}\over \mu_{\rs A}} \ {1\over \eta} \ ,
	\nonumber \\
	r_{\rs M}^2  = {f\over \ell g} \ {1\over \eta_{\rs M}} = {f\over \ell g} \sqrt{\mu_A\over \mu_{\rs N}} {1\over \eta} \ ,
	\label{semi}
\end{eqnarray}
with 
\begin{equation}
	\eta= \sqrt{\eta_{\rs Q} \eta_{\rs M}}=2\pi (mv\ell) G/\sqrt{\mu_{\rs N} \mu_{\rs A}} \ . 
	\label{etasingle}
\end{equation}
Of course, configurations with $Q\ne 0$ and $M\ne 0$ must be excluded since the two types of excitations are different, only pairs $\{ 0, M\}$ or $\{Q, 0\}$ have to be considered. The phase diagram is found by establishing which integer charges lie within the ellipse when the semi-axes are varied. This yields Eq.\,(\ref{phasestructure}) in the main text. 
\\

\subsection{Electromagnetic response in the magnetic monopole condensate} 
To establish the electromagnetic response of the monopole condensate we add the minimal coupling of the charge current $j^{\mu}$ to the electromagnetic field,
\begin{equation}
	S \to S - i \sum_x \ell^4 A_{\mu} j_{\mu} = S - i\sum_x \ell^4 {1\over 4\pi} A_{\mu} k_{\mu \alpha \beta} b_{\alpha \beta} \ ,
	\label{mincou}
\end{equation} 
and we compute its effective action by integrating over the fictitious gauge fields $a_{\mu}$ and $b_{\mu \nu}$. This requires no new computation since, by a summation by parts, the above coupling amounts only to a shift
\begin{equation}
	m_{\mu \nu} \to m_{\mu \nu} - {1\over 2\pi} \ell^2 \hat k_{\mu \nu \alpha} A_{\alpha} \ ,
	\label{shift}
\end{equation}
in (\ref{topeffac}). 
Setting $q_{\mu} = 0$ for the phase with gapped Cooper pairs gives Eq.\,(\ref{villain}) in the main text. 
\\

\subsection{Computation of the string tension}
The starting point is equation (\ref{wilson1}) in the main text. For large values of the coupling $f$, the action is peaked around the values $F_{\mu \nu}=2\pi t_{\mu \nu}$, allowing for the saddle-point approximation to compute the Wilson loop. Using the lattice Stoke's theorem, one rewrites Eq.\,(\ref{wilson1}) as
\begin{equation}
	\langle W(C) \rangle = {1\over Z_{A_{\mu}, m_{\mu \nu}}} \sum_{\{ m_{\mu \nu} \} }  \int_{-\pi}^{+\pi}  {\cal D} A_{\mu} 
	\ {\rm e}^{-{1\over 8 f^2}  \ \sum_x \left( \tilde F_{\mu \nu} -2\pi m_{\mu \nu}  \right)^2 } {\rm e}^{i{q_{\rm ext} \over 2}\sum_S S_{\mu \nu} (\tilde F_{\mu \nu} -2\pi m_{\mu \nu}) } \ ,
	\label{wilson2}
\end{equation}
where the quantities $S_{\mu \nu}$ are unit surface elements perpendicular (in 4D) to the plaquettes forming the surface $S$ encircled by the loop $C$ and vanish on all other plaquettes. We have also multiplied the Wilson loop operator by 1 in the form ${\rm exp} (-i \pi q_{\rm ext} \sum_x S_{\mu \nu} m_{\mu \nu})$. Following Polyakov\,\cite{polyakov}, we decompose $m_{\mu \nu}$ into transverse and longitudinal components, 
\begin{eqnarray}
	&&m_{\mu \nu} = {m^{\rm T}}_{\mu \nu}+ {m^{\rm L}}_{\mu \nu} \ ,
	\nonumber \\
	&&{m^{\rm T}}_{\mu \nu} = \ell \hat k_{\mu \nu \alpha} n_{\alpha} + \ell \hat k_{\mu \nu \alpha} \xi_{\alpha}  \ ,
	\nonumber \\ 
	&&{m^{\rm L}}_{\mu \nu} =\Delta_{\mu} \lambda_{\nu} - \Delta_{\nu} \lambda_{\mu} \ , 
	\label{decomp}
\end{eqnarray} 
where $\{ n_{\mu} \} $ are integers and we adopt the gauge choice $\Delta_{\mu} \lambda_{\mu} = 0$, so that 
$\nabla^2 \lambda_{\mu} =\hat \Delta_{\nu} \Delta_{\nu} \lambda_{\mu} = m_{\mu}$, with $m_{\mu} \in {\mathbb Z} $ describe the world-lines of the magnetic monopoles on the lattice. The set of 6 integers $\{ m_{\mu \nu}\} $ has thus been traded for 3 integers $\{ n_{\mu} \} $ and 3 integers $\{ m_{\mu} \}$ representing the magnetic monopoles. The former are then used to shift the integration domain for the gauge field $A_{\mu}$ to $[-\infty,+\infty]$. The real variables $\{ \xi_{\mu} \}$ can then also be absorbed into the gauge field. The integral over the now non-compact gauge field $A_{\mu}$ gives the Gaussian fluctuations around the saddle points $m_{\mu}$. Gaussian fluctuations contribute the usual Coulomb potential $1/|x|$ in 3D. We shall henceforth focus only on the magnetic monopoles.
\begin{equation}
	\langle W(C) \rangle = {1\over Z_{ m_{\mu}}} \sum_{\{ m_{\mu} \} }  \ e^{-{\pi^2 \over 2 f^2} \sum_{x, \mu} m_{\mu} 
		{1\over -\nabla^2} m_{\mu}} \ e^{ i2\pi q_{\rm ext} \sum_S m_{\mu} {1\over -\nabla^2}  \hat \Delta_{\nu} S_{\nu \mu} } \ .
	\label{wilson3}
\end{equation} 

Following\,\cite{orland} we introduce a dual gauge field $\chi_{\mu}$ with field strength $g_{\mu \nu} = \Delta_{\mu} \chi_{\nu} -\Delta_{\nu} \chi_{\mu}$ and we rewrite (\ref{wilson3}) as
\begin{equation} 
	\langle W(C) \rangle = {1\over Z_{ m_{\mu}, \chi_{\mu}}} \int {\cal D} \chi_{\mu} \ {\rm e}^{-{f^2\over 2\pi^2 }\sum_{x, \mu, \nu} g_{\mu \nu}^2}
	\sum_{N} {z^N\over N!} \sum_{x_1,\dots , x_N}  \sum_{m^1_{\mu},\dots , m^n_{\mu} = \pm1} e^{i \sum_{x, \mu} m_{\mu} ( \chi_{\mu} + q_{\rm ext} \eta_{\mu})} \ ,
	\label{chi}
\end{equation}
where the angle $\eta_{\mu} = 2\pi \hat \Delta_{\nu} S_{\nu \mu}/(-\nabla^2)$ represents a dipole sheet on the Wilson surface $S$ and the monopole fugacity $z$ is determined by the self-interaction as
\begin{equation}
	z = e^{-{\pi^2 \over 2f^2} G(0) } \ ,
	\label{fugacity}
\end{equation}
with $G(0)$ being the inverse of the Laplacian at coinciding arguments. We also used the dilute gas approximation, valid at large $f$, in which one takes into account only single monopoles $m_{\mu}=\pm 1$. The sum can now be explicitly performed\,\cite{orland}, with the result,
\begin{equation}
	\langle W(C) \rangle = {1\over Z_{ \chi_{\mu}}} \int {\cal D} \chi_{\mu} \ e^{-{f^2\over 2\pi^2}\sum_{x,\mu, \nu} g_{\mu \nu}^2
		+ {4\pi^2\over f^2 } z (1- {\rm cos}(\chi_{\mu} + q_{\rm ext}\eta_{\mu}  ))} \ ,
	\label{sinegordon1}
\end{equation}
By shifting the gauge field $\chi_{\mu}$ by $-q_{\rm ext}\eta_{\mu}$ and introducing $M^2 =( \pi^2 / 2f^2) z  $, we can rewrite this as
\begin{equation}
	\langle W(C) \rangle = {1\over Z_{\chi_{\mu}}} \int {\cal D} \chi_{\mu} \ e^{-{2f^2\over \pi^2}\sum_{x,\mu,\nu} {1\over 4} {g^{\prime}_{\mu \nu}}^2 + M^2 (1- {\rm cos}(\chi_{\mu} ))} \ ,
	\label{sinegordon2}
\end{equation}
where $g^{\prime}_{\mu \nu} = g_{\mu \nu} \left( \chi_{\mu} -q_{\rm ext} \eta_{\mu} \right)$. 
For large $f$, this integral is dominated by the classical solution to the equation of motion
\begin{equation}
	\hat \Delta_{\mu} g_{\mu \nu}^{\rm cl} = - 2\pi q_{\rm ext} \hat \Delta_{\mu} S_{\mu \nu} + M^2 {\rm sin} \chi_{\nu}^{\rm cl} \ .
	\label{eqmo1}
\end{equation}

Let us assume that the Wilson loop lies in the (0-3) plane formed by the Euclidean time direction 0 and the z axis. In this case, there are non-trivial solutions only for the 1- and 2-components of the gauge field, while $\chi_3^{\rm cl} = 0$. With the Ansatz $\chi_1^{\rm cl} = \chi_1^{\rm cl} (x_2)$, $\chi_2^{\rm cl} = \chi_2^{\rm cl} (x_1)$, we are left with two one-dimensional equations in the region far from the boundaries of the Wilson surface $S$, 
\begin{eqnarray}
	\hat \Delta_{1} \Delta_{1} \chi_2^{\rm cl} = - 2\pi q_{\rm ext} \hat \Delta_{1} S_{12} + M^2 {\rm sin} \chi_2^{\rm cl} \ ,
	\nonumber \\
	\hat \Delta_{2} \Delta_{2} \chi_1^{\rm cl} = - 2\pi q_{\rm ext} \hat \Delta_{2} S_{21} + M^2 {\rm sin} \chi_1^{\rm cl} \ ,
	\label{eqmo2}
\end{eqnarray}
Following\,\cite{polyakov}, we solve these equations in the continuum limit,
\begin{eqnarray}
	\partial_{1} \partial_{1} \chi_2^{\rm cl} = 2\pi q_{\rm ext} \delta^{\prime} (x_1) +M^2 {\rm sin} \chi_2^{\rm cl} \ ,
	\nonumber \\
	\partial_{2} \partial_{2} \chi_1^{\rm cl} = 2\pi q_{\rm ext} \delta^{\prime} (x_2) +M^2 {\rm sin} \chi_1^{\rm cl} \ .
	\label{eqmo3}
\end{eqnarray}
For $q_{\rm ext} = 1$ (corresponding to Cooper pairs in our case), the classical solutions with the boundary conditions $\chi_{1,2}^{\rm cl} \to 0$ for $|x_{1,2}| \to \infty$ are 
\begin{eqnarray}
	\chi_1^{\rm cl} = {\rm sign} (x_2) \ 4 \ {\rm arctan} \  {\rm e}^{-M |x_2|} \ ,
	\nonumber \\
	\chi_2^{\rm cl} = {\rm sign} (x_1) \ 4 \ {\rm arctan} \  {\rm e}^{-M |x_1|} \ .
	\label{cooperpairs}
\end{eqnarray}
Inserting this back in (\ref{sinegordon2}) we get formula (\ref{arealaw}) in the main text.





\vspace{-0.2cm}



\vspace{-0.2cm}


\begin{thebibliography}{10}



\bibitem{olive}
Goddard,\,P. \& Olive,\,D.\,I.  Magnetic monopoles in gauge field theories. {\it Rep. Prog. Phys.} {\bf 41},  1357--1437 (1978).

\bibitem{milton}
Milton,\,K.\,A.  Theoretical and experimental status of magnetic monopoles. {\it Rep. Prog. Phys.} {\bf 69} 1637-1712 (2006). 


\bibitem{qi} 
Qi,\,X.\,L., Li,\, R., Zhang,\,J. \& Zhang,\,S.-C. Inducing a magnetic monopole with topological surface states. \textit{Science} {\bf 323}, 1184--1187 (2009).


\bibitem{zeldov} 
Uri,\,A. \textit{et al}. Nanoscale imaging of equilibrium quantum Hall edge currents and of the magnetic monopole response in graphene. {\it Nature Physics} {\bf 16}, 164--170 (2020). 

\bibitem{dirac1931}
Dirac,\,P.\,A.\,M. Quantised singularities in the electromagnetic field. \textit{Proc. R. Soc}\,A\,\textbf{133}, 60--72 (1931).

\bibitem{thooft1974}
't Hooft,\,G. Magnetic Monopoles in Unified Gauge Theories. \textit{Nucl. Phys.} B\,\textbf{79}, 267--284 (1974).

\bibitem{polyakov1974}
Polyakov,\,A.\,M. Particle spectrum in quantum field theory. \textit{JETP Lett}. \textbf{20}, 194-195 (1974).

\bibitem{nambu}Nambu,\,Y. Strings, monopoles and gauge fields. {\it Phys. Rev.} {\bf D10} 4262 (1974).

\bibitem{mandelstam}Mandelstam,\, S. Vortices and quark confinement in non-Abelian gauge theories. {\it Phys. Rep.} {\bf 23C} 
245-249 (1976). 

\bibitem{thooft}'t Hooft,\, G. On the phase transition towards permanent quark confinement. {\it Nucl. Phys.} {\bf B138} 1346-1349 (1978). 

\bibitem{Blanter1997}
Ya.\,M.\,Blanter, R.\,Fazio, and G.\,Sch\"on. Duality in Josephson Junction Arrays. \textit{Nuclear Physics}\,B\,(Proc. Suppl.) \textbf{58}, 79 -- 90 (1997).

\bibitem{dst}
Diamantini,\,M.\,C., Sodano,\,P. \& Trugenberger,\,C.\,A.  Gauge theories of Josephson junction arrays. \textit{Nuclear Physics} B\textbf{474}, 641 -- 677 (1996).

\bibitem{vinokur}
Vinokur,\,V.\,M.  \textit{et~al.} Superinsulator and quantum synchronization. \textit{Nature} \textbf{452}, 613 -- 615 (2008). 

\bibitem{dtv1}
Diamantini,\,M.\,C., Trugenberger,\,C.\,A. \& Vinokur,\,V.\,M. Confinement and asymptotic freedom with Cooper pairs. \textit{Comm. Phys.} \textbf{1}, 77 (2018).

\bibitem{dtnpv}
Trugenberger,\,C., Diamantini,\,M.\,C.,  Nogueira,\,F.\,S., Poccia,\,N. \& Vinokur,\,V.\,M. Magnetic Monopoles and Superinsulation in Josephson Junction Arrays. \textit{Quantum Reports} \textbf{2}, 388 -- 399 (2020).

\bibitem{Tinkham}
Tinkham,\,M. Introduction to superconductivity. McGraw-Hill, Inc. 1996.





\bibitem{mironov}
Mironov,\,A.\,Yu. \textit{et al}.
Charge Berezinskii-Kosterlitz-Thouless transition in superconducting NbTiN films. {\it Scientific Reports} {\bf 8} 4082 (2018). 


\bibitem{dtv2}
Diamantini,\,M.\,C., Gammaitoni,\,L., Strunk,\,C., Postolova,\,S.\,V.,  Mironov,\,A\,Yu.,  Trugenberger,\,C.\,A. \& Vinokur,\,V.\,M.  Direct probe of the interior of an electric pion in a Cooper pair superinsulator. \textit{Communications Physics} \textbf{3}, 142 (2020). 

\bibitem{greensite}
Greensite,\,J. {\it An introduction to the confinement problem.} Springer-Verlag (Berlin) 2011. 


\bibitem{Kowal1994}
Kowal,\,D. \& Ovadyahu,\,Z. Disorder induced granularity in an amorphous superconductor. \textit{Solid St. Comm.} \textbf{90},783 -- 786 (1994).

\bibitem{Fistul2008}
Fistul,\,M.\,V., Vinokur,\,V.\,M.\, \&Baturina,\,T.\.I.
Collective Cooper-Pair Transport in the Insulating State of Josephson-Junction Arrays. \textit{Phys. Rev. Lett.} \textbf{100}, 086805 (2008).

\bibitem{Vinokur2013}
Baturina,\,T.\,I. \& Vinokur,\,V.\,M. Superinsulator–superconductor duality in two dimensions. \textit{Annals of Physics}, \textbf{331}, 236 -- 257 (2013).

\bibitem{dtv2019}
Diamantini,\,M.\,C., Trugenberger,\,C.\,A. \& Vinokur,\,V.\,M. Topological gauge theory of the superconductor-insulator transition. In \textit{Topological Phase Transitions and New Developments, pp 197 -- 221, World Scientific, 2019}.

\bibitem{pancake1} 
Blatter,\,G., Feigel'man,\,M.\,V., Geshkenbein,\,V.\,B., Larkin,\,A.\,I.  \& Vinokur,\,V.\,M.  Vortices in high-temperature supercondcutors. {\it Rev. Mod. Phys.} {\bf 66}, 1125--1388 (1994). 

\bibitem{wilczek}
Wilczek,\,F., Disassembling Anyons. {\it Phys. Rev. Lett.} {\bf 69} 132-135 (1992).

\bibitem{blau}
Birmingham,\,D., Blau,\, M., Rakowski,\,M. \& Thompson,\,G. Topological field theory {\it Phys. Rep.} {\bf 209}, 129--340 (1991).

\bibitem{wen2013}
Wen,\,X.-G. Topological Order: From Long-Range Entangled Quantum Matter to a Unified Origin of Light and Electrons. \textit{ISRN} \textbf{2013}, 198710 (2013). https://doi.org/10.1155/2013/198710 

\bibitem{semenoff}
Bergeron,\,M., Semenoff\,G.\,W. \& Szabo,\,R. Canonical BF-type topological field theory and fractional statistics of strings. {\it Nucl. Phys.} B\,{\bf 437}, 695--722 (1995).

\bibitem{senthil} 
Vishwanath,\,A. \& Senthil,\,T. Physics of Three-Dimensional Bosonic Topological Insulators: Surface-Deconfined Criticality and Quantized Magnetoelectric Effect. {\it Phys. Rev. } X\,{\bf 3}, 011016 (2013).






\bibitem{moschalkov}
Zhang,\,G. \textit{et. al}. Metal-bosonic insulator-superconductor transition in boron-doped granular diamond. {\it Phys. Rev. Lett} {\bf 110},  077001 (2013). 

\bibitem{jja}Fazio,\,R. and van\,der\,Zant,\,H. Quantum phase transitions and vortex dynamics in superconducting networks.
\textit{Physics Reports}  {\bf 355}, 235 -- 334 (2001). 

\bibitem{bowick}
Allen,\,T.,  Bowick,\,M. \& Lahiri,\,A.  Topological Mass Generation in 3+1 dimensions. {\it Mod. Phys. Lett.} A\,{\bf 6},  559--571 (1991).

\bibitem{onsager1}
Onsager,\,L. Statistical hydrodynamics {\it Nuovo Cimento Supp.}  {\bf 6}, 279--287 (1949).



\bibitem{feynman}
Feynman,\,R. {\it Statistical Mechanics} Benjamin, Reading (1972). 

\bibitem{schakel} 
Schakel,\,A.\,M.\,J.  Percolation, Bose-Einstein condensation and string proliferation. {\it Phys. Rev. } E\,{\bf 63}, 026115 (2001). 

\bibitem{kleinert}
Kleinert,\,H.  \& Chervyakov,\,A. Evidence for negative stiffness of QCD flux tubes in the large-N limit of SU(N). {\it Physics Letters} B\,{\bf 381}, 286--290 (1996). 


\bibitem{geostring}
Diamantini,\,M.\,C. \& Trugenberger,\,C.\,A.  Geometric aspects of confining strings. {\it Nucl. Phys.} B\,{\bf 531}, 151--167 (1998). 

\bibitem{diamantini2020}
Diamantini,\,M.\,C. \textit{et al}. Bosonic topological insulator intermediate state in the superconductor-insulator transition. \textit{Physics Letters}\,A\,\textbf{384}, 126570 (2020).

\bibitem{qcd} Schmidt,\, C. \& Sharma,\,S. The phase structure of QCD. {\it Jour. Phys. G: Nucl. Part. Phys.} {\bf 44} 104002 (2017). 

\bibitem{polyakovoriginal}
Polyakov,\,A. Compact gauge fields and the infrared catastrophe. {\it Physics Letters} B\,{\bf 59}, 82--84 (1975). 

\bibitem{polyakov}
Polyakov,\,A.\,M.  {\it Gauge Fields and Strings}, Harwood Academic Publisher, Chur (Switzerland) (1987). 

\bibitem{dtv4}
Diamantini,\,M.\,C. \textit{et al}. Bosonic topological insulator intermediate state in the superconductor-insulator transition. \textit{Physics Letters} A\,\textbf{384}, 126570 (2020).	

\bibitem{Zhou2019}
Zhou,\,P. \textit{et al}, Electron pairing in the pseudogap state revealed by
shot noise in copper oxide junctions. \textit{Nature} \textbf{572}, 493--496 (2019).

\bibitem{dtv3} 	
Diamantini,\,M.\,C., Gammaitoni,\,L., Trugenberger,\,C.\,A.  \& Vinokur,\,V.\,M.  Voger-Fulcher-Tamman criticality of 3D superinsulators, {\it Scientific Reports} {\bf 8}, 15718 (2018).

\bibitem{shahar2005}
Sambandamurthy,\, G., Engel,\,L.\,M., Johansson,\, A., Peled,\,E. \& Shahar,\,D. Experimental evidence for a collective insulating state in two-dimensional superconductors.\textit{Phys. Rev. Lett}. \textbf{94}, 017003 (2005).

\bibitem{ovadia2015}
Ovadia,\,M. \textit{et al}. Evidence for a finite-temperature insulator. \textit{Sci. Rep}. \textbf{5}, 13503 (2015).



\bibitem{nelson} Nelson,\,D., Piran,\,T., \& Weinberg,\,S. eds. {\it Statistical Mechanics of Membranes and Surfaces}, World Scientific, Singapors (2004). 

\bibitem{orland} Orland,\,P. Instantons and disorder in antisymmetric tensor gauge fields. {\it Nucl. Phys.} B\,{\bf 205}, 107-118 (1982). 








\end{thebibliography}
\end{document}